\documentclass[twocolumn]{aastex631}

\shorttitle{Jet--RTN mapping and non-gravitational accelerations of 3I/ATLAS}
\shortauthors{Scarmato and Loeb}

\usepackage{amsmath,amssymb}
\usepackage{graphicx}
\usepackage{hyperref}

\newcommand{\Rhat}{\hat{\mathbf R}}
\newcommand{\That}{\hat{\mathbf T}}
\newcommand{\Nhat}{\hat{\mathbf N}}
\newcommand{\lhat}{\hat{\boldsymbol\ell}}

\begin{document}

\title{Linking System of Jets to the Non-Gravitational Acceleration of 3I/ATLAS}\

\author{Toni Scarmato}
\affiliation{Toni Scarmato's Astronomical Observatory, San Costantino di Briatico, Calabria, 89817, Italy}

\author{Abraham Loeb}
\affiliation{Astronomy Department, Harvard University, 60 Garden Street, Cambridge MA 02138, USA}

\begin{abstract}
Building on the jet morphology and periodic wobble analysis of 3I/ATLAS in \citet{ScarmatoLoeb2026}, we link observed jet position angles (PAs) and the non-gravitational acceleration components $(A_1,A_2,A_3)$ in the 3D RTN (radial, transverse, normal) frame relative to the Sun. We: (i) compute RTN directions from heliocentric state vectors and project them on the sky at the measured astrometric pointings; (ii) compare projected RTN PAs to three persistent jets (Jet1-Jet2-Jet3) and quantify angular offsets; and (iii) estimate order-of-magnitude thrust and accelerations from HST/WFC3-UVIS F350LP net counts via transformations from photometry to cross-section, dust mass, mass-loss rate, and thrust. We explicitly document the uncertainties through background handling, phase-function systematics, and geometric degeneracies along the line of sight. For U.T. 2025-11-30.80903, Jet2 is aligned with the projected transverse direction to within $\sim0.5^\circ$, while Jet3 is the closest to the projected normal direction with a moderate offset ($\sim25^\circ$). For UT 2025-12-27, Jet2 exhibits a monotonic PA drift over $\sim24$ minutes with a larger oscillation amplitude. The results provide a reproducible RTN-projection workflow and compact jet--RTN correspondence tables suitable for non-gravitational acceleration modeling of 3I/ATLAS.
\end{abstract}


\section{Introduction}
The interstellar object 3I/ATLAS showed persistent inner-coma jet structures, with evolving position angles (PAs) and multiple periodicities, as reported in \citet{ScarmatoLoeb2026}. In cometary dynamics, non-gravitational accelerations are commonly expressed in the RTN (radial, transverse, normal) frame relative to the Sun as
\begin{equation}
a_{\rm NG}(t) = A_1 \hat{R} + A_2 \hat{T} + A_3 \hat{N}
\end{equation}
where $\Rhat$ is the unit vector radial relative to the Sun, $\That$ is transverse along the nucleus trajectory and $\Nhat$ is normal to the orbital plane. Here we link the observed jet directions to these components using sky-plane RTN projections and provide order-of-magnitude thrust estimates from HST photometry. This manuscript  follows on the results reported in \citet{ScarmatoLoeb2026}, based on Hubble Space Telescope (HST) images available at \href{https://doi.org/10.17909/g3j5-kq58}{MAST}.
\\
\\
\section{Data and Inputs}
\subsection{HST observations and jet PAs}
We adopted the observed jet PAs in the sky (with N=0$^\circ$, E=90$^\circ$) for UT 2025-11-30.80903:
\begin{itemize}
\item Jet1: $PA=65^\circ$,
\item Jet2: $PA=290^\circ$,
\item Jet3: $PA=175^\circ$.
\end{itemize}
At UT 2025-12-27.68194, Jet2 is measured at $PA \simeq 255^\circ$ and shows a short-timescale drift in a sequence:
263$^\circ$, 260$^\circ$, 258$^\circ$, 256$^\circ$, 255$^\circ$ (each $\pm 3^\circ$).
As reported in \citet{ScarmatoLoeb2026}, Jet2 oscillates within roughly $\pm20^\circ$ around $\sim270^\circ$.

\subsection{Astrometry and auxiliary angles}
For UT 2025-11-30.80903 we use:
\begin{itemize}
\item Pointing: RA=12:05:18, Dec=+00:08:52 (J2000),
\item Solar PA: $PA_\odot = 113^\circ$ ($PA_{\rm anti\text{-}Sun}=293^\circ$),
\item Apparent motion rates: $\dot{\alpha} = -2.18''\,{\rm h}^{-1}$,
$\dot{\delta} = +0.77''\,{\rm h}^{-1}$,
where $\dot{\alpha}$ and $\dot{\delta}$ denote the apparent
rates of motion in right ascension and declination,
respectively, projected on the sky plane. (motion PA $\approx 289^\circ$).
\end{itemize}
For UT 2025-12-27 (image sequence) we use:
\begin{itemize}
\item Pointing: RA=10:09:02, Dec=+10:07:03 (J2000),
\item Solar PA: $PA_\odot = 105^\circ$ ($PA_{\rm anti\text{-}Sun}=285^\circ$).
\end{itemize}

\subsection{HST photometric calibration and dynamics assumptions}
We adopt WFC3/UVIS drc parameters as used in the associated calculations:
\begin{itemize}
\item Filter: F350LP, exposure time $t_{\rm exp}=270$ s,
\item Gain: 1.5 e$^-$/DN,
\item PHOTFLAM: $5.24350573\times10^{-20}$ erg s$^{-1}$ cm$^{-2}$ \AA$^{-1}$ per (e$^-$/s),
\item PHOTPLAM: 5851.1484 \AA,
\item Pixel scale: 0.04$''$/px, aperture radius: 5 px (0.20$''$).
\end{itemize}
For thrust estimates we assume:
\begin{itemize}
\item Dust speed $v_d=120$ m/s, gas speed $v_g=400$ m/s, gas-to-dust mass-loss ratio $\chi=1$,
\item Nucleus radius $R_N=1.3$ km and bulk density $\rho_N=500$ kg m$^{-3}$,
\item Dust bulk density $\rho_d=1000$ kg m$^{-3}$ and effective grain size $a_{\rm eff}=10\,\mu$m (results scale linearly with $a_{\rm eff}$),
\item Dust phase function: Halley--Marcus (HM) at $\alpha\simeq 65^\circ$ (see Section~\ref{sec:phase}).
\end{itemize}
We emphasize that these are order-of-magnitude assumptions; the resulting accelerations carry factor is of few systematics, dominated by dust phase function and size distribution.
The adopted parameters for the thrust estimates (dust velocity $v_d$, gas velocity $v_g$, 
gas-to-dust ratio $\chi$, grain size $a_{\rm eff}$, and densities) are chosen to be 
representative of typical cometary conditions reported in the literature. 

In particular, the adopted dust and gas velocities are consistent with values 
commonly inferred for moderately active comets at heliocentric distances of a few AU, 
while the assumed grain size and densities correspond to standard order-of-magnitude 
estimates used in dust coma modeling.

We emphasize that these parameters are not uniquely constrained for 3I/ATLAS, 
and the resulting thrust and acceleration estimates should be interpreted as 
order-of-magnitude values. The derived quantities scale linearly with the assumed 
grain size and depend inversely on the adopted phase function and albedo.(e.g., A’Hearn et al. 1995; Jewitt \& Meech 1987; Combi et al. 2004)
\section{Methods}
\subsection{RTN construction and sky-plane projection}
Given heliocentric state vectors $(\mathbf r,\mathbf v)$, we define
\begin{equation}
\Rhat=\frac{\mathbf r}{|\mathbf r|},\qquad
\Nhat=\frac{\mathbf r\times\mathbf v}{|\mathbf r\times\mathbf v|},\qquad
\That=\Nhat\times\Rhat.
\end{equation}
For a given observer pointing (RA, Dec), the line-of-sight unit vector is $\lhat$. We project any unit vector $\hat{\mathbf u}$ to the plane of the sky via
\begin{equation}
\hat{\mathbf u}_\perp=\hat{\mathbf u}-(\hat{\mathbf u}\cdot\lhat)\lhat.
\end{equation}
The sky-plane position angle $PA(\hat{\mathbf u})$ (N=0$^\circ$, E=90$^\circ$) is then computed from $\hat{\mathbf u}_\perp$ expressed in the local tangent basis (North, East). A critical diagnostic is the projected norm $\|\hat{\mathbf u}_\perp\|$: if $\|\Rhat_\perp\|\ll1$, then the radial direction is nearly along the line of sight and $PA(\Rhat)$ is not diagnostically constrained.

\subsection{Photometry thrust pipeline}
For each jet region, the net DN sum (assumed already background/coma-subtracted) is converted to a count-rate
\begin{equation}
\dot N = \frac{\mathrm{DN}_{\rm net}\,(\mathrm{GAIN})}{t_{\rm exp}} \quad [\mathrm{e^-\,s^{-1}}],
\end{equation}
and to a flux density at the pivot wavelength
\begin{equation}
F_\lambda = \dot N\;\mathrm{PHOTFLAM}.
\end{equation}
DN$_{\rm net}$ represents the background-subtracted detector signal in data 
numbers within the selected aperture. The detector gain converts data numbers 
to electrons via $N_{\rm e^-} = {\rm DN} \times {\rm GAIN}$, allowing the 
determination of the count rate $\dot{N}$ in electrons per second.

The conversion from count rate to physical flux density is performed using 
the PHOTFLAM calibration keyword, which provides the inverse sensitivity 
(i.e., the flux density corresponding to a count rate of 1 e$^-$ s$^{-1}$). 
This procedure follows the standard HST/WFC3 photometric calibration 
pipeline (Dressel 2022), Wide Field Camera 3 Instrument Handbook, Version 14.0 
(Baltimore: STScI).
We estimate the dust scattering cross-section $C$ (at order-of-magnitude) via a reflected-sunlight approximation:
\begin{equation}
C \propto \frac{r^2\Delta^2}{p\,\Phi(\alpha)}\,10^{-0.4(m-m_\odot)},
\end{equation}
where $p$ is geometric albedo, $\Phi(\alpha)$ is the dust phase function, and $(r,\Delta,\alpha)$ are heliocentric distance, observer distance, and phase angle. 

The proportionality in Eq. (6) reflects the use of an order-of-magnitude 
reflected-light approximation. The absolute normalization depends on the 
adopted solar magnitude and filter bandpass, and therefore introduces an 
additional systematic uncertainty of order unity in the derived cross-section.

For a single adopted grain size,
\begin{equation}
M_d \approx \frac{4}{3}\rho_d\,a_{\rm eff}\,C.
\end{equation}
Assuming that the material leaves the observed aperture over a residence time $t_{\rm res}\approx \rho/v_d$, we estimate
\begin{equation}
\dot m_d \approx \frac{M_d}{t_{\rm res}}.
\end{equation}
Including gas momentum with ratio $\chi=\dot m_g/\dot m_d$, the thrust is approximated as
\begin{equation}
F \approx \dot m_d\,(v_d+\chi v_g),
\end{equation}
and the resulting acceleration magnitude is
\begin{equation}
|a| \approx \frac{F}{M_N}, \qquad
M_N=\frac{4\pi}{3}R_N^3\rho_N.
\end{equation}
Equation (9) approximates the combined dust and gas momentum contribution.
\subsection{Phase-function systematics at $\alpha\sim 65^\circ$}
\label{sec:phase}
A key inconsistency encountered is the dust phase function choice at large phase angles. A simplistic linear law $\Phi_{\rm lin}(\alpha)=10^{-0.4\beta\alpha}$ (with $\beta\simeq0.04$ mag/deg) yields $\Phi_{\rm lin}\approx0.09$ at $\alpha\simeq65^\circ$, whereas the Halley--Marcus (HM) composite phase function tabulated by the Lowell Observatory \footnote{https://asteroid.lowell.edu/comet/dustphase/table} gives $\Phi_{\rm HM}\approx0.356$ at $\alpha\simeq65^\circ$. Because $C$, $M_d$, thrust, and $|a|$ scale as $1/\Phi$, adopting HM reduces these quantities by a factor $\Phi_{\rm lin}/\Phi_{\rm HM}\approx0.26$ relative to the linear law.
The HM composite dust phase function is adopted to account for 
the angular dependence of light scattering by cometary dust. This empirical 
function, derived from observations of comet 1P/Halley and subsequent datasets, 
provides a more realistic description of dust scattering compared to a simple 
linear phase law, particularly at intermediate phase angles ($\alpha \sim 60^\circ$--$70^\circ$).

At the phase angle relevant to this work ($\alpha \simeq 65^\circ$), the HM phase 
function yields $\Phi_{\rm HM} \approx 0.356$, significantly larger than the value 
$\Phi_{\rm lin} \approx 0.09$ obtained from a linear phase law with $\beta \sim 0.04$ mag deg$^{-1}$. 
Since the derived dust cross-section, mass, and thrust scale as $1/\Phi(\alpha)$, 
the choice of phase function introduces a systematic factor of $\sim 4$ in the 
resulting physical quantities.

We therefore adopt the HM phase function as a more physically representative 
description of cometary dust scattering, while noting that phase-function 
uncertainties remain one of the dominant sources of systematic error (e.g., Schleicher et al. 1998; Marcus 2007; Schleicher \& Bair 2011; 
see also Lowell Observatory dust phase function documentation).
We note that the choice of phase function represents one of the dominant 
sources of systematic uncertainty, as all derived quantities scale as 
$\Phi^{-1}(\alpha)$. Alternative phase-function prescriptions would therefore 
shift the inferred cross-sections, masses, and accelerations by factors of a few.

\section{Results}
\subsection{RTN sky-plane PA comparison (UT 2025-11-30.80903)}
Using the heliocentric state vector at JD 2461010.30903 and the pointing RA=12:05:18, Dec = +00:08:52, we obtained:
\begin{equation}
PA(\That)=290^\circ,\qquad PA(\Nhat)=200^\circ,
\end{equation}
while $\|\Rhat_\perp\|\approx 1.9\times10^{-4}$, implying that $\Rhat$ is nearly along the line of sight and $PA(\Rhat)$ is not robust. With $PA_\odot=113^\circ$ the antisolar PA is $293^\circ$, nearly coincident with $PA(\That)$ in projection. Consequently, single-epoch PA matching cannot uniquely separate $A_1$ vs $A_2$ in this geometry, but the observed alignment is consistent with a significant contribution of A2, 
although a residual degeneracy with the radial component cannot be excluded 
given the near line-of-sight orientation of $\hat{R}$.

\subsection{RTN sky-plane PA comparison (UT 2025-12-27)}
For JD 2461036.5 and pointing RA=10:09:02, Dec = + 10:07:03, we obtained the following.
\begin{equation}
PA(\That)=287.6^\circ,\qquad PA(\Nhat)=197.6^\circ,
\end{equation}
and again $\|\Rhat_\perp\|\sim 10^{-4}$, implying that $PA(\Rhat)$ is not diagnostically constrained. With $PA_\odot=105^\circ$, the antisolar PA is $285^\circ$.

\subsection{Compact jet table: PA offsets and order-of-magnitude accelerations}
Table~\ref{tab:compact} summarizes jet PAs, their angular offsets relative to the antisolar direction (A1 proxy), projected transverse (A2) and projected normal (A3), together with the order-of-magnitude accelerations derived from the photometry-to-thrust pipeline under the adopted assumptions.

\begin{deluxetable*}{lrrrrrrrrrrl}
\tablecaption{Compact 2D jet--RTN correspondence and acceleration estimates (HM phase function, $a_{\rm eff}=10\,\mu$m). \label{tab:compact}}
\tablewidth{0pt}
\tablehead{
\colhead{Jet} & \colhead{DN$_{\rm net}$} & \colhead{PA} &
\colhead{$\Delta A1$} & \colhead{$\Delta A2$} & \colhead{$\Delta A3$} &
\colhead{$C$} & \colhead{$M_d$} & \colhead{$\dot m_d$} & \colhead{$F$} & \colhead{$a$} & \colhead{Best match}\\
\colhead{} & \colhead{} & \colhead{(deg)} &
\colhead{(deg)} & \colhead{(deg)} & \colhead{(deg)} &
\colhead{(km$^2$)} & \colhead{(kg)} & \colhead{(kg/s)} & \colhead{(N)} & \colhead{(m/s$^2$)} & \colhead{}
}
\startdata
Jet1 & 2.86e5 & 65  & 132   & 135.5 & 134.5 & 9.66  & 1.29e5 & 5.51e1 & 2.87e4 & 6.21e-9 & --- \\
Jet2 & 7.98e5 & 290 & 3     & 0.5   & 90.5  & 26.99 & 3.59e5 & 1.54e2 & 8.01e4 & 1.74e-8 & A1,A2 (R,T) \\
Jet3 & 1.12e5 & 175 & 118   & 114.5 & 24.5  & 3.78  & 5.04e4 & 2.16e1 & 1.12e4 & 2.44e-9 & A3 (weak) \\
\enddata
\tablecomments{Angles are computed for UT 2025-11-30.80903 using $PA_{\rm anti\text{-}Sun}=293^\circ$, $PA(\That)=289.54^\circ$, $PA(\Nhat)=199.54^\circ$. The radial direction is near the line of sight ($\|\Rhat_\perp\|\ll1$), so A1 cannot be uniquely validated by PA matching at this geometry. Acceleration estimates assume net DN already background/coma-subtracted and scale linearly with $a_{\rm eff}$ and inversely with $p\,\Phi(\alpha)$.}
\end{deluxetable*}

\subsection{Documented inconsistencies and their resolution}
We encountered and resolved the following issues during the analysis:
\begin{enumerate}
\item \textbf{Background treatment (DN):} Early calculations subtracted an additional background term even when DN values were already provided as net jet counts, artificially reducing fluxes and accelerations. The final tables treat jet DN values as net (no further subtraction).
\item \textbf{Phase function at $\alpha\simeq65^\circ$:} Switching from the linear phase law to the Halley--Marcus composite function reduces $C$, $M_d$, $F$, and $a$ by a factor $\sim0.26$ at $\alpha\simeq65^\circ$. We adopt HM for consistency.
\item \textbf{A1 vs A2 degeneracy in sky-plane:} At both epochs examined, $\|\Rhat_\perp\|\sim10^{-4}$--$10^{-3}$, so the projected radial direction is nearly undefined. This makes A1 difficult to constrain from sky-plane PAs and can cause the antisolar direction to appear nearly coincident with $\That$ in projection.
\item \textbf{Jet2 non-stationarity:} Multi-epoch PA differences for Jet2 are consistent with a wobble envelope ($\pm 20^\circ$ around the mean PA), so a single PA should not be treated as a fixed inertial direction. RTN association should consider the oscillation envelope and/or a rotation-averaged thrust direction.
\end{enumerate}

\section{Discussion: Toward a 3D (A1,A2,A3) decomposition}
A full 3D decomposition requires the jet unit vector $\hat{\mathbf j}(t)$, not only its sky-plane PA. Each PA provides a 3D plane constraint containing the line of sight; multiple epochs can recover a 3D cone/axis model for $\hat{\mathbf j}(t)$. Once $\hat{\mathbf j}(t)$ is estimated, the RTN components follow from dot products:
\begin{equation}
A_1 \propto \hat{\mathbf j}\cdot\Rhat,\qquad
A_2 \propto \hat{\mathbf j}\cdot\That,\qquad
A_3 \propto \hat{\mathbf j}\cdot\Nhat.
\end{equation}
The present 2D results already provide strong evidence that Jet2 contributes substantially to $A_2$ (transverse) at 2025-11-30, while Jet3 is only a weak/moderate candidate for $A_3$ based on PA proximity to $\Nhat$.
In this context, the inferred alignment between jet directions and RTN 
components should be interpreted as a time-averaged effect, reflecting 
both rotational modulation and projection effects, rather than a strictly 
fixed forcing direction.

\section{Jet periodicities and frequency relations}
\label{sec:freq}

\begin{deluxetable*}{lrrrrl}
\tablecaption{Jet periods and derived frequencies, including sum/difference combinations.}
\tablewidth{0pt}
\tablehead{
\colhead{Entry} &
\colhead{$P$ (hr)} &
\colhead{$f=1/P$ (hr$^{-1}$)} &
\colhead{Combination} &
\colhead{$P_{\rm equiv}$ (hr)} &
\colhead{Comment}
}
\startdata
Jet1 & 2.9 & 0.3448 & --- & --- & observed \\
Jet2 & 7.1 & 0.1408 & --- & --- & dominant wobble period \\
Jet3 & 4.9 & 0.2041 & --- & --- & secondary period \\
$f_{J2}+f_{J3}$ & --- & 0.3449 & $f_{J2}+f_{J3}$ & 2.90 & $\approx P_{J1}$ (sum sideband) \\
$|f_{J3}-f_{J2}|$ & --- & 0.0632 & $|f_{J3}-f_{J2}|$ & 15.81 & predicts $\sim$15.8 hr (difference sideband) \\
\enddata
\tablecomments{Frequencies are computed from the jet PA modulation periods reported in \citet{ScarmatoLoeb2026}.}
\end{deluxetable*}

\citet{ScarmatoLoeb2026} reports three distinct jet PA modulation periods, which we denote
$P_{J1}$, $P_{J2}$, and $P_{J3}$ for Jets 1--3, respectively. Using the values
$P_{J1}=2.9$~h, $P_{J2}=7.1$~h, and $P_{J3}=4.9$~h, the corresponding frequencies are
\begin{equation}
f_k=\frac{1}{P_k}\qquad [\mathrm{h}^{-1}].
\end{equation}
Numerically,
\begin{equation}
f_{J1}=0.35,\quad
f_{J2}=0.14,\quad
f_{J3}=0.20\ \mathrm{hr}^{-1}.
\end{equation}
A noteworthy relation is that the Jet~1 frequency is consistent with the sum of the Jet~2 and Jet~3 frequencies:
\begin{equation}
f_{J1}\simeq f_{J2}+f_{J3} \quad \Rightarrow \quad
\frac{1}{2.9\ \mathrm{hr}}\simeq \frac{1}{7.1\ \mathrm{hr}}+\frac{1}{4.9\ \mathrm{hr}},
\end{equation}
since
\begin{equation}
f_{J2}+f_{J3}=0.35\ \mathrm{hr}^{-1}
\quad \Leftrightarrow \quad
P_{\rm sum}=\frac{1}{f_{J2}+f_{J3}}=2.90\ \mathrm{hr},
\end{equation}
which matches $P_{J1}$ within rounding. This type of linear frequency combination is naturally produced when the
jet morphology is modulated by more than one physical timescale (e.g., non-principal-axis rotation with two
fundamental modes producing sum/difference sidebands), or by geometric/visibility effects that mix two underlying
periodicities. In the RTN framework developed here, such multi-period modulation implies that a single-epoch PA
cannot be treated as a fixed inertial jet direction; instead, the relevant forcing direction should be considered
as a time-dependent cone (or an average over rotational phase), while the magnitude of the forcing can be tracked
via the photometry-to-thrust pipeline.

These results suggest that the observed jet activity may be linked 
to a complex rotational state of the nucleus, potentially involving 
non-principal-axis rotation or time-variable active regions.
This relation should be regarded as suggestive rather than conclusive, 
as a detailed dynamical model would be required to establish a firm 
physical interpretation of the observed frequency combinations.
\\
\\
\\
\\
\section{Conclusions}
In this paper, we provided: (i) a reproducible RTN sky-projection framework, (ii) a compact jet--RTN correspondence table, and (iii) order-of-magnitude non-gravitational accelerations derived from HST photometry under stated assumptions, while documenting the principal sources of uncertainty and degeneracy. The analysis supports Jet2 as the dominant contributor to transverse non-gravitational acceleration, forcing A2 in the examined geometry, with A1 generally not diagnosable from PA when $\Rhat$ is nearly along the line of sight. Future work should fit a 3D cone model for each jet using multi-epoch PA sequences to obtain robust $(A_1,A_2,A_3)$ decompositions.
\\

{ACKNOWLEDGMENTS}\\
{A.L. was supported in part by the Harvard Black Hole Initiative (funded by GBMF and JTF) and  Galileo Project. This work was carried out using Astroart for image processing, Astrometrica for time photometry, open-source Python tools for time-series analysis and figure generation, including Astropy, Numpy, Scipy and MatPlotlib. The authors thank NASA, ESA, and STScI for the HST data used in this work.}

\begin{figure*}[t]
\centering
\includegraphics[width=0.95\textwidth]{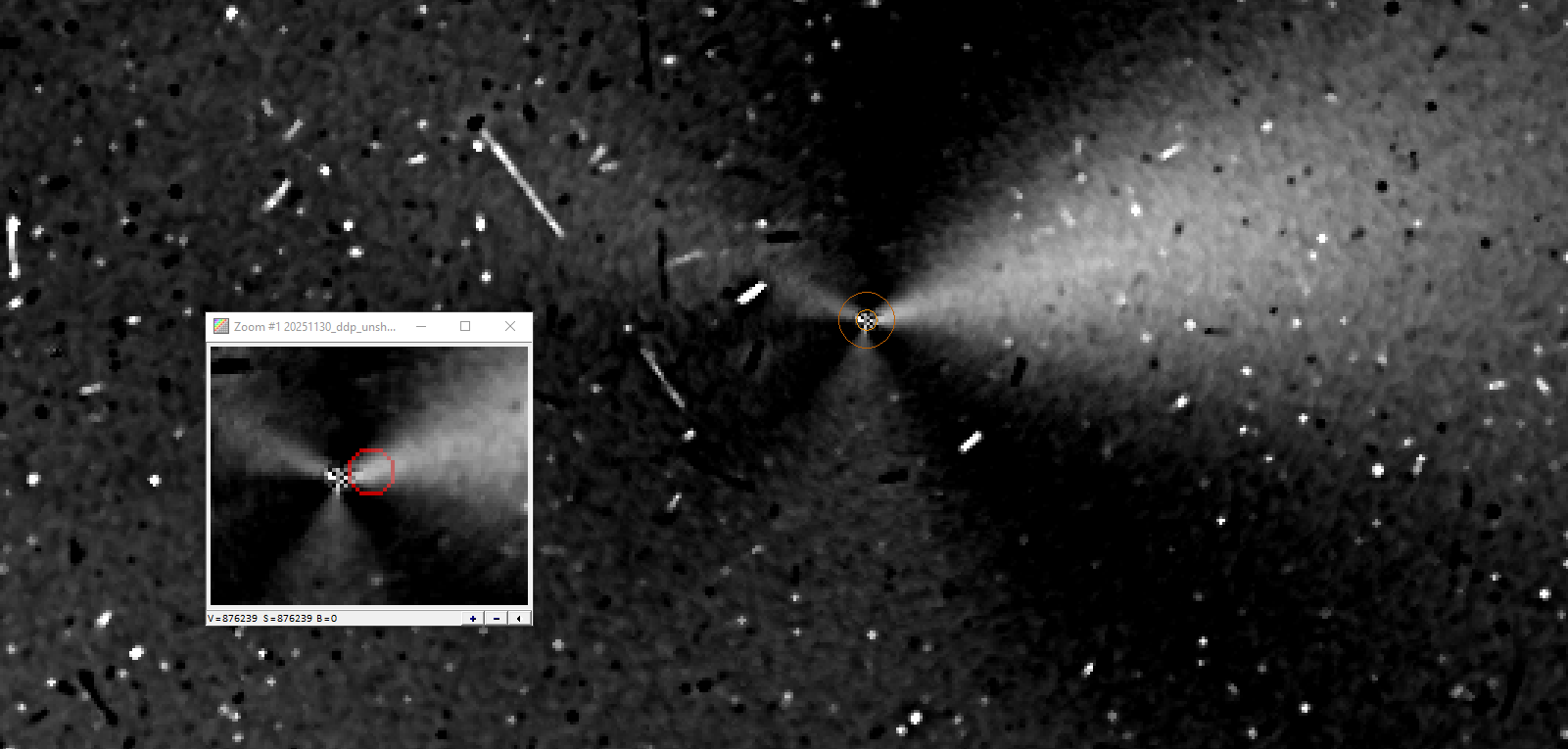}
\caption{Reference inner-coma morphology showing persistent jet structures 
(illustrative). North is up and East is to the left. The RTN-mapping 
procedure uses the measured jet PAs and the astrometric pointing to 
compare with projected RTN directions computed from state vectors.}
\end{figure*}

\end{document}